\documentclass[10pt, journal]{IEEEtran}

\ifCLASSINFOpdf
\else
   \usepackage[dvips]{graphicx}
\fi
\usepackage{url}
\usepackage{amsmath,amsfonts,amssymb,amsthm}

\usepackage[numbers]{natbib}
\usepackage{graphicx}

\begin{document}

\title{HiFi-Stream: Streaming Speech Enhancement with Generative Adversarial Networks}

\author{Ekaterina Dmitrieva, and Maksim Kaledin.
\thanks{E. Dmitrieva is with HSE University, Russia, Moscow, 109028 11 Pokrovsky Bulvar (email: evdmitrieva\_3@edu.hse.ru). }
\thanks{M. Kaledin is with HSE University, Russia, Moscow, 109028 11 Pokrovsky Bulvar (email: mkaledin@hse.ru). Corresponding author. }
}

\markboth{Preprint}
{Shell \MakeLowercase{\textit{et al.}}: Made with IEEEtran.cls for IEEE Journals}
\maketitle

\begin{abstract}
Speech Enhancement techniques have become core technologies in mobile devices and voice software. Still, modern deep learning solutions often require high amount of computational resources what makes their usage on low-resource devices challenging. We present HiFi-Stream, an optimized version of recently published HiFi++ model. Our experiments demonstrate that HiFi-Stream saves most of the qualities of the original model despite its size and computational complexity improved in comparison to the original HiFi++ making it one of the smallest and fastest models available. The model is evaluated in streaming setting  where it demonstrates its superior performance in comparison to modern baselines.
\end{abstract}



\section{Introduction}
Speech enhancement (SE) techniques are generally used as a preprocessing tool for speech downstream tasks such as speech recognition \cite{ wang:VFLitev2:2021,wang:VFLite:2020}. The main goal of SE is to remove noise and reverberations coming from the actual environment where the audio is recorded. In the area of SE, the state-of-the-art Deep Learning (DL) approaches demonstrated that DL can manage very hard audio conditions in SE and source separation problems. Typically DL models require powerful computational units which are unlikely to be placed in a mobile or wearable device. Construction of architectures for speech processing tasks better suited for low-resource applications has recently become the focus of research efforts \cite{wu:TrueLowPower:2024}.


In our work we propose HiFi-Stream models, two optimized versions of HiFi++ \cite{andreev:hifiPP:2023} introducing structural and technical modifications. The experimental set-up, pipeline and model configurations are given in the github repository: \url{https://github.com/KVDmitrieva/source_sep_hifi}.
\begin{enumerate}
    \item The experiments on VCTK \cite{valentinibotinhao16_interspeech} dataset in offline setting demonstrate that the proposed model does not lose much in performance in comparison to the original model.
    \item HiFi-Stream models are reduced at least nearly 40\% in parameters and around 10\% in MACs making them a viable quality-complexity trade-offs.
    \item We evaluate the trained HiFi-Stream models in audio stream simulation where the audio is processed in chunks. HiFi-Stream outperforms its direct competitors: HiFi++\cite{andreev:hifiPP:2023} and FFC-SE\cite{shchekotov:ffcse:22}; without additional streaming training. We found the new attention-like module for mask construction to be the key ingredient resulting in model robustness in limited context.
\end{enumerate}

\begin{figure*}
    \centering
    \includegraphics[width=1\linewidth]{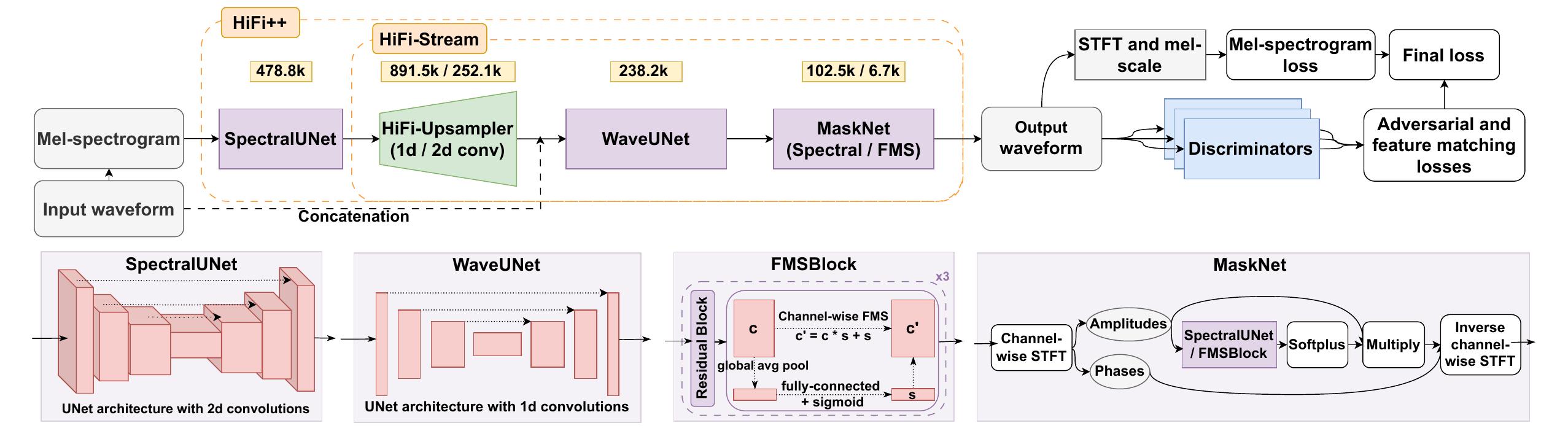}
    \caption{HiFi++ and Hifi-Stream architectures: the new model reduced the number of parameters, got new MaskNet block, redesigned HiFi upsampler and removed Spectral U-Net.}
    \label{fig:hifiScheme}
\end{figure*}

\section{Related Work}
SE and speech separation methods generally could be classified into time-frequency (TF) \cite{ chen:fullsubnetplus:2022, hu:dccrn:2020, wang:VFLitev2:2021} and time-domain (TD) \cite{ defossez:demucsSE:2020, luo:dprnn:2020,tstnn:2021} models. Despite the growing interest in time-domain models, they remain to be very complex with at least millions of parameters \cite{luo:dprnn:2020}. Studies demonstrate that TF methods have advantage over TD methods because magnitude spectrogram serves as very expressive and in the same time is processed very efficiently \cite{chen:fullsubnetplus:2022, hu:dccrn:2020}.

Streaming processing presents several specific constraints \cite{wu:TrueLowPower:2024}. The first is computational complexity reflected by the number of parameters and processing time. The second is causality: the model is restricted to use only limited context. So, the researchers are focused on constructing streaming-aware versions of state-of-the-art solutions based on transformer architectures \cite{ chenda:skim:2022}. It was recently shown that transformer-based solutions can be implemented in low-resource setting but with considerable efforts invested into the attention mechanisms \cite{wu:TrueLowPower:2024}.

The generative look at the problem presented by SEGAN \cite{pascual:segan:2017} later developed into HiFi-GAN solutions \cite{su:hifigan1:2020,su:hifigan2:2021}. It was suggested to train and design the model as GAN where the generator predicts the clean waveform. This was reviewed in HiFi++ \cite{andreev:hifiPP:2023} with more structured design. The model achieved the baseline results despite significantly reduced model complexity. Several other solutions of the same scale were proposed later \cite{ liu:iccrn:2023, shchekotov:ffcse:22, zhao:sicrn:2024}. 
In \cite{andreev:hifiPP:2023}, \cite{shchekotov:ffcse:22} the main goal was to reduce the model size but computational complexity was not addressed. Our aim is to improve also the computational complexity of the SE, and we improved the results of \cite{andreev:hifiPP:2023,shchekotov:ffcse:22} in this regard, see Fig. \ref{fig:models}.

\section{Methodology}

The problem of speech enhancement is to get an estimate of the clean speech signal $x \in \mathbb{R}^{N_{ch} \times T}$ with $N_{ch}$ channels from its corrupted representation
\begin{equation}
y = f \star x + \varepsilon,
\end{equation}
where $f \in \mathbb{R}^{N_{ch} \times k}$ represents an impulse response (IR) kernel (the effects of the environment acoustics, reverberation and recording device) and $\star$ depicts a convolution operation. In such notation, the IR kernel is convolved with the clean signal and $\varepsilon \in \mathbb{R}^{N_{ch} \times T}$ serves as an additive noise component. In our work we consider a problem with $N_{ch}=1$ though the modification for other cases can be done straightforwardly. Originally GAN approaches mostly used time-domain approach requiring direct signal reconstruction and implying large number of parameters since one needs quite complex encoder and decoder to obtain a representative set of features from raw audio. So, we in our work take a route of TF-methods rising from original vocoding applications. Classic TF approaches usually propose to estimate a magnitude mask $M(y) \in \mathbb{R}^{F \times M}$ with values from $0$ to $1$ for signal spectrogram $s = STFT(y) \in \mathbb{C}^{F \times M}$ and then estimate the clean signal as
\begin{equation}
\hat{x} = iSTFT(M(y) \odot s)
\end{equation}
with $\odot$ denoting coordinate-wise product. In such a way, the phase component is bypassed to have a well-defined inverse STFT. At a high level, in HiFi++, the signal is first re-synthesized across several audio channels and then processed by the masking pipeline, which produces the spectral mask.

The generator depicted on the Fig. \ref{fig:hifiScheme} consists of several blocks: STFT-Mel transform, Spectral U-Net, HiFi, Wave U-Net and SpectralMaskNet. The idea is to train the generator to predict clean waveform while discriminator decides whether the audio is clean. The model operates in steps.
\paragraph{Spectral U-Net} The block is responsible for mel-spectrogram pre-processing  and acts as feature transformer highlighting the important parts of the mel-spectrogram. We show that for mild noisy conditions the block is irrelevant, it is in line with the findings of \cite{andreev:hifiPP:2023} where the same was demonstrated for SE problem. So, we have removed the block in all our considered architectures resulting in reduction of 300 k parameters.
\paragraph{Improved HiFi block} HiFi block is the largest core block in the model responsible for upsampling the magnitude spectrogram back to signal domain. When trained, it transforms mel-spectrogram $m=Mel(s)$ of the signal $x$ into multichannel audio $x' \in \mathbb{R}^{n_{ch} \times T}$. The channels can be interpreted as signal and noise layers very similar to those observed in the output of source separation. We experimented with using 2D convolutions across time and frequency dimensions instead of 1D (across frequency only) and varying kernel size and channel parameters, which resulted in a drastic reduction of nearly 400k parameters since the number of out channels in convolutions reduced around 4 to 8 times. However, at the same time, GMACs slightly increased because of higher computational complexity of 2D convolutions. Still, we increased the performance, since the model better exploits the frequency-temporal correlations in the spectrogram.

\paragraph{Wave U-Net} is the main post-processing unit responsible for fixing the signal in time domain by removing robo-voice artifacts typical for vocoding \cite{hifigan}. With the help of input waveform concatenated with multichannel HiFi output it effectively fixes the phase shifts. The Wave U-Net design is taken from \cite{andreev:hifiPP:2023}.

\begin{figure}
    \centering
    \includegraphics[scale=0.7]{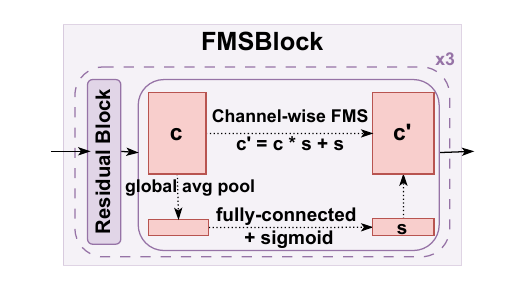}
    \caption{FMS Block inspired by \cite{trueFMS} takes the input tensor $C \in \mathbb{R}^{F \times T}$, computes the vector of scaling coefficients $S \in \mathbb{R}^{T}$ and applies them across filters and computes the result tensor $C'_{ft} = C_{ft}S_{t} + C_{ft}$. This method turns out to be a key component of model robustness in streaming mode and serves as temporal attention.}
    \label{fig:fms}
\end{figure}

\paragraph{Spectral Mask Net FMS} SpectralMaskNet is another key component in the model since it generates the spectral magnitude mask $M(y)$. HiFi++ used a U-Net convolutional processing unit. It outputs mask of gains, essentialy a tensor of the same size as spectrogram. Such processing is very natural for attention architectures. However, despite allowing for more expressive models, attention architecture is computationally expensive for streaming \cite{wu:TrueLowPower:2024}. We came up with a cheaper attention alternative from RawNet2 \cite{trueFMS} called Feature Map Scaling (FMS), known also as Squeeze-Excitation \cite{sqEx2018}. For masking we constructed FMS Blocks with affine aggregation (see Fig. \ref{fig:fms}) in a ResNet structure. FMS acts as more complexity-friendly attention alternative. This choice resulted in 20x reduction in MaskNet block in comparison to the original U-Net because of less convolution channels. In new MaskNet with just $(8,16,8)$ channels with FMS transform in between we achieved better computational efficiency due to the attention nature of FMS and absence of the bottleneck. Our key finding is that attention-like mechanism showed its robustness in the streaming setting with reduced context (see Table \ref{tab:streamOtherSE}).


\begin{table*}[ht!]
\centering
\caption{Results on VCTK \cite{valentinibotinhao16_interspeech} dataset for different modifications of HiFi++ and other SE models (the results marked with $^*$ are taken from the papers, otherwise the validation was ours). The top-2 values in top and bottom parts are put in bold.}
    \label{tab:otherSE}
    \begin{tabular}{|l|ccccccc||cc|}
    \hline
         Model  & SI-SDR (dB) $\uparrow$ & SDR (dB) $\uparrow$ & STOI $\uparrow$ & PESQ $\uparrow$ & CSIG $\uparrow$ & CBAK $\uparrow$ & COVL $\uparrow$ & nParams & GMACs/sec \\
         \hline
         \hline
         Ground Truth&  - & - & 1.000 & 4.640 & 5 & 5 & 5 & - & -\\
         Input&  8.440 & 8.54 & 0.790 & 1.970 & 3.340 & 2.820 & 2.740 & - & - \\
         \hline
         SEGAN\cite{pascual:segan:2017} &  - & - & - & 2.160 & 3.480 & 2.940 & 2.80 & - & - \\
         MetricGAN+\cite{fu:metricganplus:2021} &  8.162 & 14.734 & 0.853 & \textbf{3.130} & 3.971 & 3.123 & 3.549 & 2.700M & 28.500  \\
         DEMUCS(non-caus.)\cite{defossez:demucsSE:2020} &  \textbf{18.585} & \textbf{19.603} & \textbf{0.893} & 3.015 & 4.348 & \textbf{3.597} & \textbf{3.715} & 30.100M & 38.1  \\
         HiFiGAN-SE* \cite{su:hifigan1:2020}  &  - & - & - & 2.940  & 4.070 & 3.070 & 3.490 & - & -  \\
         HiFiGAN-2* \cite{su:hifigan2:2021}  &  - & - & - & \textbf{3.110} & \textbf{4.370} & 3.540 & 3.740 & - & -  \\
         FFC-AE-V0 \cite{shchekotov:ffcse:22} &  16.313 & 17.291 & 0.865 & 2.669 & 4.061 & 3.296 & 3.373 & \textbf{0.422M} & \textbf{4.064}  \\
         FFC-AE-V1 \cite{shchekotov:ffcse:22}&  16.899 & 17.484 & \textbf{0.893} & 2.868 & 4.238 & 3.497 & 3.573 & \textbf{1.665M} & 15.672  \\
         FFC-UNet \cite{shchekotov:ffcse:22}&  17.851 & 18.702 & \textbf{0.894} & 2.962 & \textbf{4.349} & \textbf{3.510} & \textbf{3.682} & 7.673M & 35.552  \\
         HiFi++ \cite{andreev:hifiPP:2023} &  \textbf{17.954} & \textbf{18.936} & 0.890 & 2.906 & 4.271 & 3.483 & 3.609 & 1.706M & \textbf{2.746}\\
         \hline
         HiFi++(reprod.)&  13.400 & 15.661 & \textbf{0.866} & 2.595 & 3.998 & \textbf{3.173} & 3.305 & 1.617M & 2.096 \\
         HiFi w/o spec&  12.808 & \textbf{16.155} & \textbf{0.873} & \textbf{2.703} & \textbf{4.094} & 3.166 & \textbf{3.413} & 1.356M & 2.007\\
         HiFi-2dMRF&  13.455 & 15.812 & 0.865 & 2.466 & 3.865 & 3.075 & 3.172 & \textbf{0.677M} & 2.084\\
         HiFi-Stream &  \textbf{14.067} & 15.924 & 0.865 & \textbf{2.701} & \textbf{4.031} & \textbf{3.220} & \textbf{3.372} & 1.174M & \textbf{1.895}\\
         HiFi-Stream2D &  \textbf{14.762} & \textbf{17.213} & 0.864 & 2.652 & 4.027 & 3.167 & 3.345 & \textbf{0.497M} & \textbf{1.973}\\
    \hline
    \end{tabular}
    
\end{table*}

\section{Experimental Setup}
\begin{table*}[t]
\centering
\caption{Streaming evaluation results on VCTK \cite{valentinibotinhao16_interspeech} dataset for different modifications of HiFi++ and other SE models. The top values in top and bottom parts are put in bold.}
    \label{tab:streamOtherSE}
    \begin{tabular}{|l|ccccccc|}
    \hline
         Model &  SI-SDR (dB) $\uparrow$ & SDR (dB) $\uparrow$ & STOI $\uparrow$ & PESQ $\uparrow$ & CSIG $\uparrow$ & CBAK $\uparrow$ & COVL $\uparrow$ \\
         \hline
         \hline
         Ground Truth&  - & - & 1.000 & 4.640 & 5 & 5 & 5 \\
         Input&  8.440 & 8.54 & 0.790 & 1.970 & 3.340 & 2.820 & 2.740 \\
         \hline
         DEMUCS(caus.dns)\cite{defossez:demucsSE:2020} &  \textbf{17.804} & \textbf{19.310} & 0.881 & \textbf{2.657} & 3.823 & \textbf{3.231} & \textbf{3.236}   \\
         FFC-AE-V0 \cite{shchekotov:ffcse:22} &  12.664 & 13.820 & 0.789 & 1.996 & 3.502 & 2.807 & 2.742   \\
         FFC-AE-V1 \cite{shchekotov:ffcse:22}&  15.147 & 15.806 & 0.860 & 2.424 & 3.906 & 3.214 & 3.172   \\
         FFC-UNet \cite{shchekotov:ffcse:22}&  15.683 & 16.467 & 0.861 & 2.452 & \textbf{3.974} & 3.225 & 3.222   \\
         HiFi++ \cite{andreev:hifiPP:2023} &  14.669 & 15.264 & \textbf{0.930} & 2.365 & 3.838 & 3.162 & 3.103 \\
         \hline
         HiFi++(reprod.)&  13.014 & 15.383 & 0.841 & 2.263 & \textbf{3.715} & 2.917 & \textbf{2.992}  \\
         HiFi w/o spec&  11.839 & 14.746 & 0.791 & 1.932 & 3.200 & 2.671 & 2.559 \\
         HiFi-2dMRF&  12.771 & 14.861 & 0.806 & 1.931 & 3.259 & 2.726 & 2.587 \\
         HiFi-Stream &  13.328 & 15.123 & 0.845 & \textbf{2.295} & 3.667 & 2.860 & 2.981 \\
         HiFi-Stream2D &  \textbf{14.189} & \textbf{16.328} & \textbf{0.847} & 2.286 & 3.680 & \textbf{2.925} & 2.982 \\
    \hline
    \end{tabular}
    
\end{table*}
\begin{figure*}[ht!]   \centering
    \includegraphics[scale=.25]{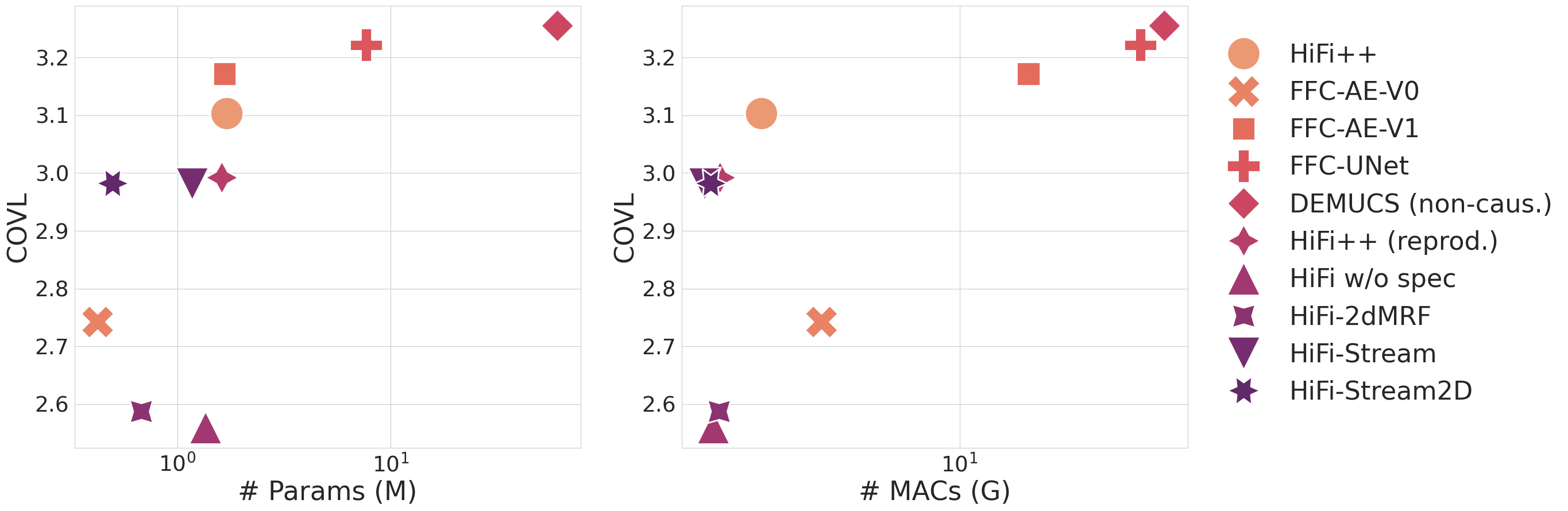}
    \caption{Performance metric (COVL) against the computational complexity and number of parameters for compared models (streaming evaluation).}
    \label{fig:models}
\end{figure*}

\subsection{Data and Evaluation}

We used Valentini VCTK \cite{valentinibotinhao16_interspeech} dataset for experiments. The dataset is a selection of $30$ speakers from the VoiceBank corpus with $28$ and $2$ speakers set for the training and test respectively. The speech utterances were mixed with noise utterances (exact specification is given in \cite{valentinibotinhao16_interspeech}). The original sample rate of the dataset is $48$ kHz, we resampled the audio to $16$ kHz for comparison with other SE solutions. All audios staged for train and validation parts were cut to $3$ seconds while test part is not changed \cite{andreev:hifiPP:2023}.

In streaming scenario we used chunks of 4096 samples with no intersections. This gives us an estimate of streaming performance and measures the effect of limited context. The model separately processes all the chunks and the processed chunks are joined together to emulate the processing pipeline in audio stream.

For evaluation we used perceptual evaluation metrics representing the quality and intelligibility of the speech: CSIG, CBAK, COVL \cite{andreev:hifiPP:2023}, WB-PESQ \cite{metrics:pesq}, STOI \cite{metrics:stoi}. Also, we present objective evaluation metrics: Signal-To-Distortion Ration (SDR, dB) 
\begin{equation}
SDR(x, \hat{x}) = 20 \log_{10} \frac{\Vert x \Vert}{\Vert x - \hat{x} \Vert }
\end{equation}
and its scale-invariant version (SI-SDR, dB) given by \cite{metrics:sisdr} to ensure the amplitude invariance.
The key aspect of our research is model complexity, so we also demonstrate the complexity metrics: number of parameters and GMACs per second of audio measured with pytorch profiling library \texttt{thop}. 

\subsection{Training details}

All models were trained for 150 epochs with learning rate 2e-4 on the given dataset with batch size 32 using AdamW optimizer (see the repository for the exact details). We used the same loss construction and discriminators as in \cite{andreev:hifiPP:2023, hifigan} and focus our main efforts on the generator structure. The generator loss is a composite loss 
\begin{equation}
\mathcal{L}(\theta) = \mathcal{L}_{GAN}(\theta) + \alpha_{FM} \mathcal{L}_{FM}(\theta) + \alpha_{Mel}\mathcal{L}_{Mel}(\theta)
\end{equation}
consisting of adversarial, feature-matching and mel reconstruction losses with $\alpha_{FM}=2$ and $\alpha_{Mel}=45$. Discriminators (3 MSD as in \cite{andreev:hifiPP:2023}) are  trained with only optimizing adversarial component
\begin{equation}
\mathcal{L}(\phi_i) = \mathcal{L}_{GAN}(\phi_i)
\end{equation}
for $i$-th discriminator. More specific details for reproduction are available in our configuration files. 

\section{Results}

\subsection{Offline Comparison (Full Audio)}

The results of the offline evaluation on the test set are shown in Table \ref{tab:otherSE}. The top part presents baseline models from the literature while the bottom part shows our own trained models.

Paper description differs from the checkpoint presented in HiFi++ repository  in the number of  convolution channels in the HiFi block. It has slightly different parameter count and complexity. Also the github checkpoint was trained for longer time and showed better metrics. We compare the modified models to our reproduction (named \textit{HiFi++(reprod)}) to have the same training conditions for fair comparison but also display the original checkpoint results (denoted as \textit{HiFi++}, top part). 

\textit{HiFi-w/o-spec} is \textit{HiFi++(reprod)} without Spectral U-Net and other modifications are built on top of this model. From the results we observe that model \textit{HiFi-2dMRF} with time-frequency 2D convolutions and reduced number of convolution channels showed drastic reduction in parameters of HiFi Block resulting in almost 3x parameter reduction in comparison to \textit{HiFi-w/o-spec}.

The effect of FMS is presented by models \textit{HiFi-Stream} and \textit{HiFi-Stream2D}.  The former is \textit{HiFi-w/o-spec} model with FMS MaskNet in the end. It allowed not only the reduced complexity and parameters but also better quality. \textit{HiFi-Stream2D} possesses additionally 2D convolutions in HiFi MRF blocks and it allowed even further parameter reduction.

As we can see from the Table \ref{tab:otherSE}, despite the reduced complexity, the modifications with FMS (\textit{HiFi-Stream} and \textit{HiFiStream2D})  demonstrate relatively the same metrics or better than \textit{HiFi w/o spec} and \textit{HiFi++(reprod.)}. 

We compared our solutions to the available baselines including some recent GAN-based light-weight solutions.  According to Table \ref{tab:otherSE}, \textit{HiFi-Stream} models demonstrate similar performance metrics in comparison to existing low-parameter baselines. In particular, the model reaches the same results as the lightest \textit{FFC-AE-V0} with almost the same number of parameters but two times reduced computational complexity.

\subsection{Streaming Comparison}

We compare the stream-aware models in the streaming conditions (see Table \ref{tab:streamOtherSE}) without additional streaming training. It can be noted that the quality declines compared to offline evaluation in all metrics which is the effect of limited audio input and context. In comparison to published baselines \textit{HiFi++} models with FMS blocks demonstrate their robustness in this regard showing slightly worse performance. It demonstrates that FMS MaskNet exploits the given context more efficiently than U-Net in the original Spectral MaskNet.\\

Overall, \textit{HiFi-Stream2D} model can be chosen as a small-size model and \textit{HiFi-Stream} is the better option in terms of computational complexity, according to the MACs. To illustrate it more, we put the results in Fig.\ref{fig:models} showing the complexity metrics against perceptual quality given by COVL. Our models \textit{HiFi-Stream} and \textit{HiFi-Stream2D} clearly stand as better complexity-quality trade-offs than considered baselines.

\section{Conclusion}
In our work we propose two modification of HiFi++ model serving as better complexity-quality trade-off. Our results indicate that the proposed methods are more robust in changing context that makes them better suited for streaming applications.

\section*{Acknowledgment}
This research was supported in part through computational resources of HPC facilities at HSE University.

\bibliography{IEEEabrv,ref}

\begin{thebibliography}{24}
\providecommand{\natexlab}[1]{#1}
\providecommand{\url}[1]{\texttt{#1}}
\expandafter\ifx\csname urlstyle\endcsname\relax
  \providecommand{\doi}[1]{doi: #1}\else
  \providecommand{\doi}{doi: \begingroup \urlstyle{rm}\Url}\fi

\bibitem[Andreev et~al.(2023)Andreev, Alanov, Ivanov, and Vetrov]{andreev:hifiPP:2023}
P.~Andreev, A.~Alanov, O.~Ivanov, and D.~Vetrov.
\newblock Hifi++: A unified framework for bandwidth extension and speech enhancement.
\newblock In \emph{ICASSP 2023 - 2023 IEEE International Conference on Acoustics, Speech and Signal Processing (ICASSP)}, pages 1--5, 2023.
\newblock \doi{10.1109/ICASSP49357.2023.10097255}.

\bibitem[Chen et~al.(2022)Chen, Wang, Tuo, Wu, Kang, and Meng]{chen:fullsubnetplus:2022}
J.~Chen, Z.~Wang, D.~Tuo, Z.~Wu, S.~Kang, and H.~Meng.
\newblock Fullsubnet+: Channel attention fullsubnet with complex spectrograms for speech enhancement.
\newblock In \emph{ICASSP 2022 - 2022 IEEE International Conference on Acoustics, Speech and Signal Processing (ICASSP)}, pages 7857--7861, 2022.
\newblock \doi{10.1109/ICASSP43922.2022.9747888}.

\bibitem[Defossez et~al.(2020)Defossez, Synnaeve, and Adi]{defossez:demucsSE:2020}
A.~Defossez, G.~Synnaeve, and Y.~Adi.
\newblock Real time speech enhancement in the waveform domain.
\newblock In \emph{Proc. Interspeech}, pages 3291--3295, 2020.

\bibitem[Fu et~al.(2021)Fu, Yu, Hsieh, Plantinga, Ravanelli, Lu, and Tsao]{fu:metricganplus:2021}
S.-W. Fu, C.~Yu, T.-A. Hsieh, P.~Plantinga, M.~Ravanelli, X.~Lu, and Y.~Tsao.
\newblock {MetricGAN+: An Improved Version of MetricGAN for Speech Enhancement}.
\newblock In \emph{Proc. Interspeech}, pages 201--205, 2021.
\newblock \doi{10.21437/Interspeech.2021-599}.

\bibitem[Hu et~al.(2018)Hu, Shen, and Sun]{sqEx2018}
J.~Hu, L.~Shen, and G.~Sun.
\newblock Squeeze-and-excitation networks.
\newblock In \emph{2018 IEEE/CVF Conference on Computer Vision and Pattern Recognition}, pages 7132--7141, 2018.
\newblock \doi{10.1109/CVPR.2018.00745}.

\bibitem[Hu et~al.(2020)Hu, Liu, Lv, Xing, Zhang, Fu, Wu, Zhang, and Xie]{hu:dccrn:2020}
Y.~Hu, Y.~Liu, S.~Lv, M.~Xing, S.~Zhang, Y.~Fu, J.~Wu, B.~Zhang, and L.~Xie.
\newblock {DCCRN: Deep Complex Convolution Recurrent Network for Phase-Aware Speech Enhancement}.
\newblock In \emph{Proc. Interspeech}, pages 2472--2476, 2020.
\newblock \doi{10.21437/Interspeech.2020-2537}.

\bibitem[Kong et~al.(2020)Kong, Kim, and Bae]{hifigan}
J.~Kong, J.~Kim, and J.~Bae.
\newblock Hifi-gan: generative adversarial networks for efficient and high fidelity speech synthesis.
\newblock In \emph{Proceedings of the 34th International Conference on Neural Information Processing Systems}, NIPS '20, Red Hook, NY, USA, 2020. Curran Associates Inc.
\newblock ISBN 9781713829546.

\bibitem[Li et~al.(2022)Li, Yang, Wang, and Qian]{chenda:skim:2022}
C.~Li, L.~Yang, W.~Wang, and Y.~Qian.
\newblock Skim: Skipping memory lstm for low-latency real-time continuous speech separation.
\newblock In \emph{ICASSP 2022 - 2022 IEEE International Conference on Acoustics, Speech and Signal Processing (ICASSP)}, pages 681--685, 2022.
\newblock \doi{10.1109/ICASSP43922.2022.9746372}.

\bibitem[Liu and Zhang(2023)]{liu:iccrn:2023}
J.~Liu and X.~Zhang.
\newblock Iccrn: Inplace cepstral convolutional recurrent neural network for monaural speech enhancement.
\newblock In \emph{ICASSP 2023 - 2023 IEEE International Conference on Acoustics, Speech and Signal Processing (ICASSP)}, pages 1--5, 2023.
\newblock \doi{10.1109/ICASSP49357.2023.10096918}.

\bibitem[Luo et~al.(2020)Luo, Chen, and Yoshioka]{luo:dprnn:2020}
Y.~Luo, Z.~Chen, and T.~Yoshioka.
\newblock Dual-path rnn: Efficient long sequence modeling for time-domain single-channel speech separation.
\newblock In \emph{ICASSP 2020 - 2020 IEEE International Conference on Acoustics, Speech and Signal Processing (ICASSP)}, pages 46--50, 2020.
\newblock \doi{10.1109/ICASSP40776.2020.9054266}.

\bibitem[Pascual et~al.(2017)Pascual, Bonafonte, and Serrà]{pascual:segan:2017}
S.~Pascual, A.~Bonafonte, and J.~Serrà.
\newblock {SEGAN: Speech Enhancement Generative Adversarial Network}.
\newblock In \emph{Proc. Interspeech}, pages 3642--3646, 2017.
\newblock \doi{10.21437/Interspeech.2017-1428}.

\bibitem[Rikhye et~al.(2021)Rikhye, Wang, Liang, He, and McGraw]{wang:VFLitev2:2021}
R.~Rikhye, Q.~Wang, Q.~Liang, Y.~He, and I.~McGraw.
\newblock Multi-user voicefilter-lite via attentive speaker embedding.
\newblock In \emph{2021 IEEE Automatic Speech Recognition and Understanding Workshop (ASRU)}, pages 275--282, 2021.
\newblock \doi{10.1109/ASRU51503.2021.9687870}.

\bibitem[Rix et~al.(2001)Rix, Beerends, Hollier, and Hekstra]{metrics:pesq}
A.~Rix, J.~Beerends, M.~Hollier, and A.~Hekstra.
\newblock Perceptual evaluation of speech quality (pesq)-a new method for speech quality assessment of telephone networks and codecs.
\newblock In \emph{2001 IEEE International Conference on Acoustics, Speech, and Signal Processing. Proceedings (Cat. No.01CH37221)}, volume~2, pages 749--752 vol.2, 2001.
\newblock \doi{10.1109/ICASSP.2001.941023}.

\bibitem[Roux et~al.(2018)Roux, Wisdom, Erdogan, and Hershey]{metrics:sisdr}
J.~L. Roux, S.~Wisdom, H.~Erdogan, and J.~R. Hershey.
\newblock Sdr – half-baked or well done?
\newblock \emph{ICASSP 2019 - 2019 IEEE International Conference on Acoustics, Speech and Signal Processing (ICASSP)}, pages 626--630, 2018.
\newblock URL \url{https://api.semanticscholar.org/CorpusID:53246666}.

\bibitem[Shchekotov et~al.(2022)Shchekotov, Andreev, Ivanov, Alanov, and Vetrov]{shchekotov:ffcse:22}
I.~Shchekotov, P.~K. Andreev, O.~Ivanov, A.~Alanov, and D.~Vetrov.
\newblock {FFC-SE: Fast Fourier Convolution for Speech Enhancement}.
\newblock In \emph{Proc. Interspeech}, pages 1188--1192, 2022.
\newblock \doi{10.21437/Interspeech.2022-603}.

\bibitem[Su et~al.(2020)Su, Jin, and Finkelstein]{su:hifigan1:2020}
J.~Su, Z.~Jin, and A.~Finkelstein.
\newblock {HiFi-GAN: High-Fidelity Denoising and Dereverberation Based on Speech Deep Features in Adversarial Networks}.
\newblock In \emph{Proc. Interspeech}, pages 4506--4510, 2020.
\newblock \doi{10.21437/Interspeech.2020-2143}.

\bibitem[Su et~al.(2021)Su, Jin, and Finkelstein]{su:hifigan2:2021}
J.~Su, Z.~Jin, and A.~Finkelstein.
\newblock {HiFi}-{GAN}-2: Studio-quality speech enhancement via generative adversarial networks conditioned on acoustic features.
\newblock In \emph{WASPAA 2021}, Oct. 2021.

\bibitem[Taal et~al.(2011)Taal, Hendriks, Heusdens, and Jensen]{metrics:stoi}
C.~H. Taal, R.~C. Hendriks, R.~Heusdens, and J.~Jensen.
\newblock An algorithm for intelligibility prediction of time–frequency weighted noisy speech.
\newblock \emph{IEEE Transactions on Audio, Speech, and Language Processing}, 19\penalty0 (7):\penalty0 2125--2136, 2011.
\newblock \doi{10.1109/TASL.2011.2114881}.

\bibitem[Valentini-Botinhao et~al.(2016)Valentini-Botinhao, Wang, Takaki, and Yamagishi]{valentinibotinhao16_interspeech}
C.~Valentini-Botinhao, X.~Wang, S.~Takaki, and J.~Yamagishi.
\newblock {Speech Enhancement for a Noise-Robust Text-to-Speech Synthesis System Using Deep Recurrent Neural Networks}.
\newblock In \emph{Proc. Interspeech}, pages 352--356, 2016.
\newblock \doi{10.21437/Interspeech.2016-159}.

\bibitem[Wang et~al.(2021)Wang, He, and Zhu]{tstnn:2021}
K.~Wang, B.~He, and W.-P. Zhu.
\newblock Tstnn: Two-stage transformer based neural network for speech enhancement in the time domain.
\newblock In \emph{ICASSP 2021 - 2021 IEEE International Conference on Acoustics, Speech and Signal Processing (ICASSP)}, pages 7098--7102, 2021.
\newblock \doi{10.1109/ICASSP39728.2021.9413740}.

\bibitem[Wang et~al.(2020)Wang, Moreno, Saglam, Wilson, Chiao, Liu, He, Li, Pelecanos, Nika, and Gruenstein]{wang:VFLite:2020}
Q.~Wang, I.~L. Moreno, M.~Saglam, K.~Wilson, A.~Chiao, R.~Liu, Y.~He, W.~Li, J.~Pelecanos, M.~Nika, and A.~Gruenstein.
\newblock {VoiceFilter-Lite: Streaming Targeted Voice Separation for On-Device Speech Recognition}.
\newblock In \emph{Proc. Interspeech}, pages 2677--2681, 2020.
\newblock \doi{10.21437/Interspeech.2020-1193}.

\bibitem[weon Jung et~al.(2020)weon Jung, bin Kim, jin Shim, ho~Kim, and Yu]{trueFMS}
J.~weon Jung, S.~bin Kim, H.~jin Shim, J.~ho~Kim, and H.-J. Yu.
\newblock Improved rawnet with feature map scaling for text-independent speaker verification using raw waveforms.
\newblock In \emph{Proc. Interspeech}, pages 1496--1500, 2020.
\newblock \doi{10.21437/Interspeech.2020-1011}.

\bibitem[Wu and Chang(2024)]{wu:TrueLowPower:2024}
C.-H. Wu and T.-S. Chang.
\newblock A low-power streaming speech enhancement accelerator for edge devices.
\newblock \emph{IEEE Open Journal of Circuits and Systems}, 5:\penalty0 128--140, 2024.
\newblock \doi{10.1109/OJCAS.2024.3387849}.

\bibitem[Zhao et~al.(2024)Zhao, He, and Zhang]{zhao:sicrn:2024}
C.~Zhao, S.~He, and X.~Zhang.
\newblock Sicrn: Advancing speech enhancement through state space model and inplace convolution techniques.
\newblock In \emph{ICASSP 2024 - 2024 IEEE International Conference on Acoustics, Speech and Signal Processing (ICASSP)}, pages 10506--10510, 2024.
\newblock \doi{10.1109/ICASSP48485.2024.10446396}.

\end{thebibliography}

\end{document}